# Observation of $e/4$ charge at ν =1/2 in GaAs

Tomer Alkalai[1,*], Emily Hajigeorgiou[2,*], Adbhut Gupta[3], Tapas Senapati[1], Priya Tiwari[1], Chia-Tse Tai[3], Siddharth Kumar Singh[3], Kirk W. Baldwin[3], Loren N. Pfeiffer[3], Mansour Shayegan[3,†], Mitali Banerjee[2,4,†] and Moty Heiblum[1,†]

[1] Department of Condensed Matter Physics, Weizmann Institute of Science, Rehovot, Israel

[2] Institute of Physics, Ecole Polytechnique Fédérale de Lausanne, Lausanne, Switzerland

[3] Department of Electrical and Computer Engineering, Princeton, USA

[4] Center for Quantum Science and Engineering, Ecole Polytechnique Fédérale de Lausanne, Lausanne, Switzerland

[*] These authors contributed equally to this work.

[†] Corresponding authors. Email: shayegan@princeton.edu (M.S), mitali.banerjee@epfl.ch (M.B), moty.heiblum@weizmann.ac.il (M.H).

## Abstract

Even-denominator fractional quantum Hall states (FQHSs) fall outside the standard Laughlin's and Jain's odd-denominator hierarchy. In this work, we study the FQHS $\nu = 1/2$ in the lowest Landau level. The state is confined within a 70 nm-wide GaAs quantum well, where the electrons exhibit a bilayer-like charge distribution. Inter-layer interactions stabilize the $\nu = 1/2$ FQHS, which is predicted to host quasiparticles with charge $e/4$ - with either Abelian or non-Abelian topological order. Here, we report on shot-noise measurements of partitioned quasiparticles at $\nu = 1/2$, where charge partitioning is generated by a unique *etch-defined* quantum point contact. Our measurements were performed on two nominally identical devices, at two independent experimental setups. Analysis of shot noise in the weak-backscattering regime in each device reveals quasiparticles with charge $e/4$. These observations provide a clear benchmark for future studies aimed at probing the topological order of the $\nu = 1/2$ FQHS and its quasiparticles' exchange statistics.

## Introduction

Fractional Quantum Hall States (FQHSs) in two-dimensional electron systems host emergent quasiparticles that carry fractional charge, obeying exchange statistics distinct from ordinary fermions or bosons.[1] The majority of experimentally observed FQHSs occur at odd-denominator filling factors and are well described by Laughlin's and Jain's hierarchy[2,3], where their quasiparticles exhibit Abelian fractional statistics[4].

Even-denominator FQHSs present a notable exception to this paradigm. These states are believed to arise from the pairing of Composite Fermions (CFs). Moreover, they are expected to host quasiparticles with even-denominator charge and, in certain realizations, a non-Abelian exchange statistics. The latter property has generated significant interest in these states, both from a fundamental perspective and because of their potential for fault-tolerant topological quantum computation[6].

The most extensively studied even-denominator FQHS is the $\nu = 5/2$ in the second Landau level (LL)[7]. Shot-noise measurements established the existence of quasiparticles with charge $e/4$ [8], and a range of thermal transport experiments were consistent with the presence of non-Abelian anyonic excitations, and a 'Particle-Hole Pfaffian' (PH-Pf) topological order[9-12].

An even-denominator FQHS has been observed and studied in the lowest LL at $\nu = 1/2$, with its wavefunction confined in a wide GaAs quantum well[13-15], or in a double-quantum-well structure[17]. In the latter, where interlayer tunneling is negligible, the origin of the $\nu = 1/2$ FQHS is generally understood[18] in terms of the 'Halperin 331' topological order[19] (an Abelian two-component phase). In contrast, in the former $\nu = 1/2$ FQHS, where a significant interlayer tunneling is present, the topological order still remains an open question, thus requiring further studies.

Early experimental studies[15,16] of the present $\nu = 1/2$ FQHS, provided evidence that the state is being consistent with a 'two-component' description. More recent experimental and theoretical studies [20-25], however, have generated renewed interest, suggesting a paired CFs description with a non-Abelian Pfaffian-like order under appropriate conditions. Moreover, the observations of anomalously-strong FQHSs fillings, $\nu = 8/17$ and $7/13$, are interpreted as *daughter states* of the $\nu = 1/2$ FQHS with a Pfaffian order [24-28].

Both the Abelian 'Halperin 331' and the non-Abelian Pfaffian FQHSs are predicted to support quasiparticles with charge $e/4$. Therefore, measurements of the quasiparticle charge cannot distinguish between the competing topological orders. Nevertheless, establishing the presence of $e/4$ quasiparticles' charge constitutes a crucial experimental benchmark. We note that direct probes of non-Abelian statistics generally rely on interferometric schemes[30,31], time-domain braiding [29], and thermal conductance measurements [9-12].

Here, we present measurements of shot noise arising from a weak partitioning of the $\nu=1/2$ edge mode by partitioning with a unique 'etched-defined' quantum point contact (QPC) in a high-mobility 2DES confined in a wide GaAs quantum well. Measurements were performed with two identical devices (A in WIS and B in EPFL), which were studied independently by two experimental groups using separate setups. Analysis of the low-frequency shot noise as a function of the injected DC current revealed tunneling of quasiparticles carrying a charge of $e/4$.

## Device and Measurement

Samples were fabricated in a modulation-doped GaAs/AlGaAs heterostructure, consisting of a GaAs quantum well 70 nm wide. The 2DEG confined in a quantum well, with its top located 300 nm below the surface, with an electron density $n = 1.45 \cdot 10^{11}$ cm$^{-2}$ and a mobility of $1.5 \cdot 10^7$ cm$^2$/Vs measured at $T$=0.3 K. The two-terminal resistance $R_{xy}$, exhibits a $\nu = 1/2$ plateau at a magnetic field $B = 12$ T, accompanied by a vanishing longitudinal resistance (Fig. 1b).

Each device consists of a single QPC patterned at the center of the Hall bar (see schematic in Fig. 1c and SEM image in the inset of Fig.1d). In contrast to the conventional GaAs devices, the QPC constriction (width equal to 1 $\mu m$) is defined by etching, that depletes the 2DEG below, followed by a deposition of metallic gates in the etched regions of the QPC (Fig. 1d inset); allowing fine tuning of the QPC's transmission. The nonlinear transmission of each QPC as a function of the side-gate voltage is shown in Fig. 1d. Multiple ohmic contacts are patterned along the perimeter of the Hall bar (Fig. 1c).

Figure 1(a) shows the confinement potential (black) and the corresponding calculated electronic charge distribution (red) for the 70 nm-wide quantum well studied here. The degree of asymmetry in the charge distribution is quantified by the relative charge imbalance Δn/n between the two sides of the quantum well. Experimentally, this asymmetry is inferred from the energy separation Δ$_{12}$ between the lowest two electric subbands, measured from low-field Shubnikov–de Haas oscillations. We measure Δ$_{12}$≃16.4 K, which corresponds to a charge imbalance of Δn/n≈17% (see Supplementary Material S4).

## Shot-Noise Analysis

The $\nu = 1/2$ FQHS, localized in the wide quantum well, is sensitive to the symmetry of the charge density distribution across the well. A deviation from a symmetric density profile has been shown to weaken the FQHS significantly. When the asymmetry exceeds a certain threshold, the $\nu = 1/2$ state evolves into a compressible Composite Fermion sea[27,28]. In devices incorporating electrostatically defined constrictions, such as QPCs, charging the QPC's gates enhances the backscattering of the impinging current (see Fig. 1d), but also transfers charge across the QW, thereby increasing the asymmetry of the local density distribution. In the high-transmission regime, where the applied QPC's gate voltage is small, the density profile inside the QPC opening remains close to that of the unperturbed bulk.

The shot-noise data is analyzed using the standard model of independent quasiparticles tunneling, which has been widely employed in previous shot-noise measurements. In this model, weak backscattering at the QPC is treated as a stochastic sequence of uncorrelated tunneling events, where each quasiparticle with charge *e\** is reflected between the counter-propagating edge modes within the QPC.

The 'zero-frequency' spectral density of current fluctuations $S_I(\omega = 0)$ is related to the quasiparticle charge by:

$$S_I(0) = 2e^*I_{in}t(1-t),$$

where $I_{in}$ is the injected DC current, and $t$ is the transmission of the edge mode in the QPC.

At finite temperature $T$, thermal fluctuations give rise to an equilibrium Johnson–Nyquist noise, $S_I(0) = 4k_BTg$, with $g$ is the edge conductance. With an increasing current, the noise evolves smoothly from 'thermal' to 'shot-noise', yielding[33]:

$$S_I(0) = 2e^*I_{in}t(1-t)\left[coth\left(\frac{e^*V}{2k_BT}\right) - \frac{2k_BT}{e^*V}\right] + 4k_BTg.$$

This approach assumes rare, independent, tunneling events, which are expected to be valid in the weak-backscattering limit. Indeed, shot-noise measurements performed at high transmission are the most reliable for extracting the quasiparticle charge.

## Results

To validate our measurement technique and analysis framework, we first performed shot-noise measurements on well-established quantum Hall states whose quasiparticle charges have been previously measured in narrow quantum wells. Figure 2 shows representative shot-noise and transmission measurements at filling factors ν = 3 and ν = 2/3. Partitioning the outer edge mode of ν = 3 (device A), the extracted charge was consistent with the electron charge $e$. For ν = 2/3 (device B), the QPC's transmission was tuned to half, on the ⅓ conductance plateau ($t = 50\%$), with the extracted noise corresponding to a well-established Fano factor, F = $2e/3$ [34-36].

Having established the reliability of our measurement and the analysis approach, we next turn to shot-noise measurements of ν = 1/2. Figures 3b & 3d show the measured spectral density $S_I(0)$, plotted as a function of the normalized bias $eV/k_BT$ (the injected DC current plotted as the top axis), for the two independently measured devices. Both QPCs were tuned: in device A, $t_1 \sim 96\%$; while device B, $t_2 \sim 93\%$. The two nonlinear transmissions are plotted in Figs. 3a & 3c.

For both transmission values, the noise exhibits the expected crossover from thermal noise at low bias to shot-noise-dominated behavior at higher bias (see Supplementary Section 2 for further discussion).

Fitting the data to the finite-temperature expression for shot noise yields an effective quasiparticle charge of $e^* = (0.250 \pm 0.013)e$ in device A and $e^* = (0.249 \pm 0.018)e$ in device B. Within the experimental uncertainty, the extracted charges are consistent in the two transmission settings and agree with the predicted value of $e/4$. Taken together, these results support the expectation that the dominant backscattering processes at ν = 1/2 involve quasiparticles carrying charge $e/4$, and that this conclusion is robust across the range of the experimental conditions spanned by the two devices.

## Discussion and Conclusions

The shot-noise measurements presented here provide direct evidence that the dominant quasiparticles at filling factor ν = 1/2 carry fractional charges of $e/4$. The consistency of the results, obtained with two nominally identical devices, measured at two QPC transmission values and current ranges, and at different electron temperatures (in two independent experimental setups), demonstrates the robustness of our observations.

We emphasize that the observation of the quasiparticle charge of the $\nu = 1/2$ FQHS is not sufficient to uniquely determine the underlying topological order, as both Abelian and non-Abelian candidate states are predicted to support excitations with the same charge. Nevertheless, establishing the quasiparticle charge represents a necessary first step and an essential experimental benchmark for more demanding probes, such as thermal transport, time-domain braiding, and interference experiments, aimed at testing the topological order of this state. The present work, therefore, provides a well-controlled reference point for future studies of the nature of the ν=1/2 FQHS.

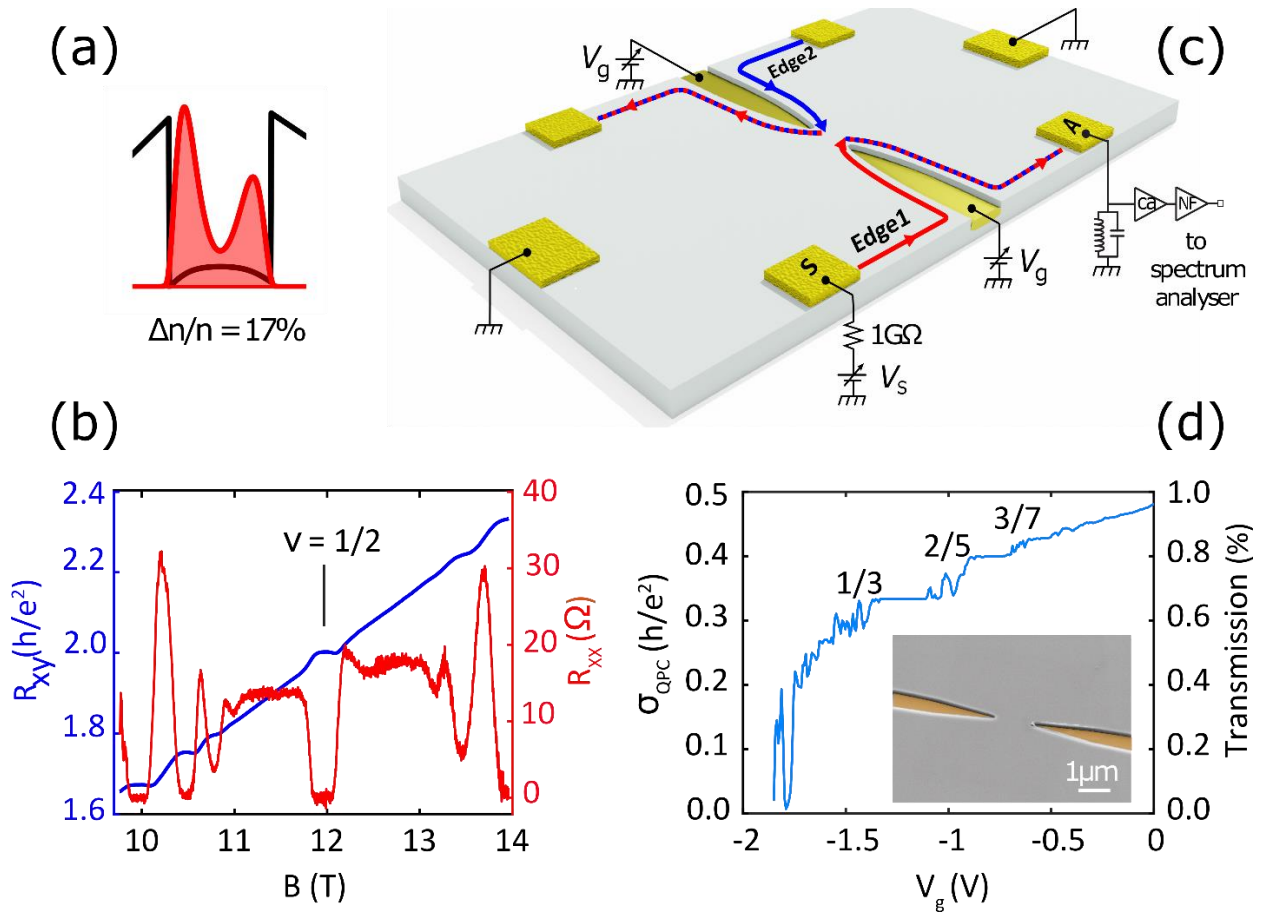

**Fig. 1:** (a) Charge distribution (red) and potential (black) for a 2DES confined to a 70nm wide quantum well. $\Delta n/n$ indicates the measured asymmetry of the charge distribution (via SDH near low $B$). (b) Hall resistance (blue trace, left axis) and longitudinal resistance (red trace, right axis) around $\nu = 1/2$. (c) A schematic of the device and the measurement circuit. A DC current is injected at the source contact (S) and propagates chirally along Edge 1. At the QPC, Edge 1 is in proximity with Edge 2, which is held at ground potential, resulting in partial current transmission and backscattering. The resulting current fluctuations (shot noise) of the transmitted edge are measured at the amplifier contact (A) (see Methods). (d) QPC conductance ($\sigma_{QPC}$, left axis) and transmission (right axis) as a function of gate voltage ($V_g$). The QPC is defined by etching, followed by a deposition of metallic gates in the etched regions. The inset shows a false-color scanning electron microscope (SEM) image of the QPC, with the metallic gates highlighted in yellow (scale bar: 1 μm). The absence of a conductance plateau at zero gate voltage arises from weak backscattering induced by the etched constriction. Tuning the gate voltage reveals conductance plateaus at $\nu_{QPC} = 1/3$, 2/5, and 3/7 (see Supplementary Section 3).

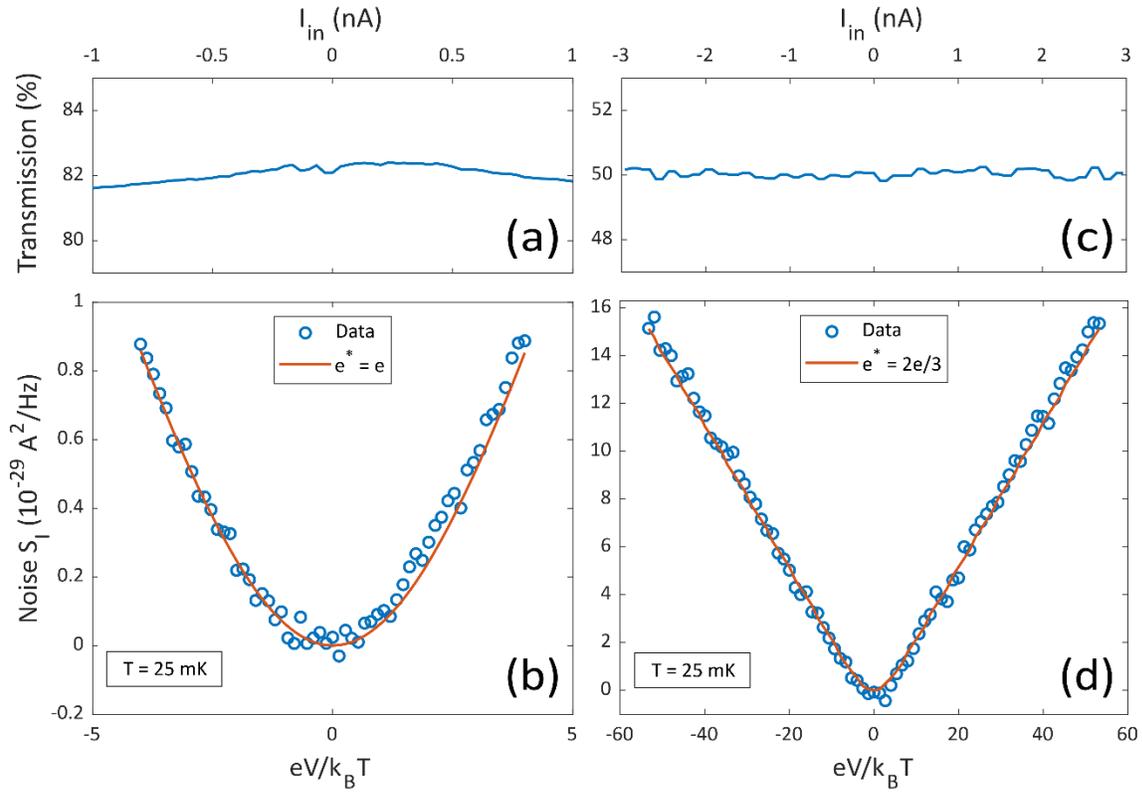

**Fig. 2:** Measured QPC transmission and spectral density of current fluctuations $S_I$ as a function of the normalized bias $eV/k_BT$. Panels (a, b) show data for $\nu = 3$ (device A), corresponding to the partitioning of the outer edge mode. Panel (a) displays the transmission, while panel (b) shows the noise spectral density; the red curve is a fit assuming quasiparticles of charge $e$, as expected. Panels (c,d) show analogous measurements for $\nu = 2/3$ (device B), taken on the 1/3 conductance plateau. Panel (c) shows the transmission, and panel (d) the noise spectral density; the red curve corresponds to an effective quasiparticle charge, $e^* = 2e/3$, consistent with previous measurements (Refs. 34-36).

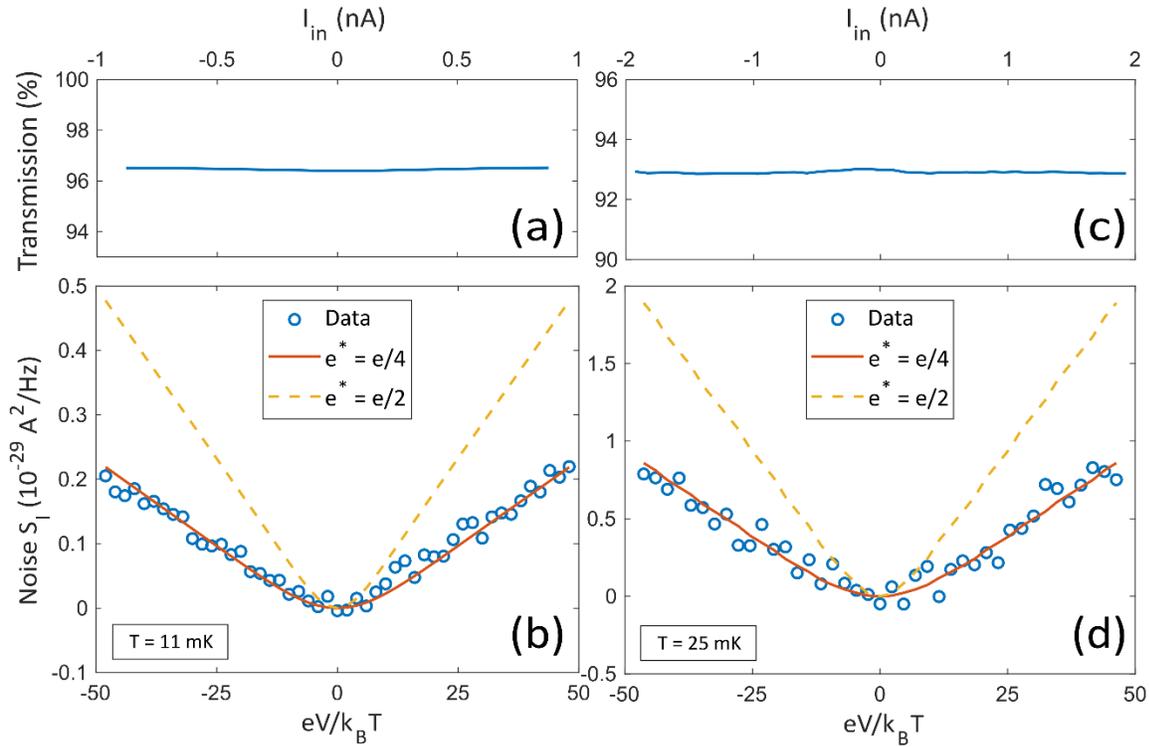

**Fig. 3:** QPC transmission and spectral density of current fluctuations $S_I$ as a function of the normalized bias $eV/k_B T$ at $\nu = 1/2$. Panels (a,b) show data for device A, measured at a transmission of 96.5%, without any voltage applied to the QPC gates, and at an electron temperature of 11 mK. Panels (c,d) show the corresponding measurements for device B, taken at a transmission of 93%, with a small applied QPC gate voltage, and at 25 mK. In both devices, the solid red curve corresponds to the expected noise for quasiparticles of charge $e^* = e/4$, while the dashed curve shows the expected noise for $e^* = e/2$ for comparison. In both panels, the data are in good agreement with the expected noise for charge $e/4$.

Data availability

The data that support the findings of this study are available from the corresponding author on request.

Author contributions:

Measurements conducted at the Weizmann Institute (device A) were performed and analyzed by T.A. , T.S. , and P.T. under the supervision of M. H. Measurements conducted at EPFL (device B) were performed and analyzed by E.H., under the supervision of M. B. The devices were fabricated by T.S. at Weizmann Institute. M.S. and S.K.S. conceived the wide well project. A.G., L.N.P., K.W.B., C.T.T., S.K.S., and M.S. designed the sample crystal structures, characterized the structures, and analyzed the data. A.G., L.N.P., and K.W.B. produced the molecular beam epitaxy samples.

All authors contributed to the write-up of the manuscript.

## Acknowledgments

For work performed in the Weizmann Institute M.H. acknowledges support of the Israel Science Foundation, Grant No. 1510/22.

For work performed at EPFL, E.H. acknowledges funding from SNSF. M.B. acknowledges support from SNSF Eccellenza grant number PCEGP2_194528, and support from the QuantERA II Programme that has received funding from the European Union's Horizon 2020 research and innovation programme under grant agreement number 101017733. M.B. also acknowledges additional financial support from EPFL SB IPHYS.

For work performed at Princeton University, we acknowledge support by the National Science Foundation (NSF) Grant No. DMR 2104771, and the Gordon and Betty Moore Foundation's EPiQS Initiative (Grant No. GBMF9615 to L. N. P.).

# Methods

1. Device Fabrication

The device (Fig. 1c) was fabricated on GaAs/AlGaAs heterostructure containing a 70 nm-wide two-dimensional electron gas (2DEG) located ∼300 nm below the surface (to the center of the well). The 2DEG has an electron density of $1.45 \times 10^{11}\,\text{cm}^{-2}$ and mobility $1.5 \cdot 10^7\,\text{cm}^2/\text{Vs}$ at $T = 0.3$ K.

The active mesa, with lateral dimensions of $450 \times 750\,\mu m^2$, was defined by wet chemical etching in an $H_2O_2$:$H_3PO_4$:$H_2O$ solution (1:1:50) for 130 s, resulting in an etch depth of approximately 200 nm. This depth exceeds the location of the Si δ-doping layer, positioned 100nm beneath the Ga(Al)As surface, thereby ensuring complete electrical isolation of the mesa. Owing to the wide quantum well's sensitivity to electron-beam exposure, optical lithography was employed for most fabrication steps, with electron-beam lithography used only once in the device process.

Ohmic contacts for current injection and voltage detection were placed 300 μm away from the quantum point contacts (QPCs). These contacts were defined using optical lithography and fabricated by electron-beam evaporation of a multilayer metal stack consisting of Ni (3 nm), Au (240 nm), Ge (130 nm), Ni (94.5 nm), and Au (20 nm), followed by rapid thermal annealing at 440 °C for 150 s.

The QPC gate patterns were defined by electron-beam lithography with a probe current of 2 nA. To enhance electrostatic coupling to the deeply buried 2DEG, the surface was dry-etched for 120 s to a depth of approximately 230 nm before gate deposition. The QPC gates were then deposited by electron-beam evaporation of Ti (5 nm) and Au (25 nm). In addition, metallic patch electrodes composed of Ti/Au (320 nm total thickness) were deposited by electron-beam evaporation. A high-magnification image of the QPC region with filled metallic gates is shown in the inset of Figure 1(d). All gate electrodes were electrically grounded during cooldown.

2. Noise acquisition setup

Device A was measured in a Leiden wet dilution refrigerator with a base temperature of 8 mK (electron temperature 11 mK), while device B was measured in a Leiden wet dilution refrigerator with a base temperature of 12 mK (electron temperature 25 mK). An LC tank circuit was connected in parallel to the sample resistance R at the contact labeled A in Fig. 1(c) of the main text, forming an RLC resonant circuit. The inductor consists of a superconducting coil for device A and a copper coil for device B, while the line capacitance provides the capacitance in both cases. The resonance frequency of the circuit is 720 kHz for device A (with a bandwidth of ∼12 kHz at filling factor ν=½) and 962 kHz for device B (with a bandwidth of ∼20kHz at ν=1/2). The RLC circuit output was amplified by a home-built cryogenic amplifier located at the 4.2 K stage, followed by additional amplification at room temperature. The amplified noise signal is analyzed using a Keysight spectrum analyzer.

3. **Thermal noise gain calibration**

The gain of the cryogenic amplifiers was calibrated using the temperature dependence of Johnson–Nyquist (JN) noise. The total voltage noise spectral density, measured at the output of the amplification chain and expressed in units of $V^2/Hz$, is given by: $S_V = G(S_{base} + S_{JN})$, where $G = G_{cold} \cdot G_{hot}$ is the total gain of the amplification chain (where $G_{hot} = 200$), $S_{base}$ is the base electronic noise of the measurement circuit, and $S_{JN} = 4k_B TR$ is the JN noise, with $R$ being the sample Hall resistance. At sufficiently high temperatures (above approximately 30–40 mK), the electronic temperature is well thermalized to the mixing-chamber temperature of $T_{MC}$. By measuring the equilibrium noise as a function of $T_{MC}$, the gain of the cryogenic amplifier $G_{cold}$ is extracted from the linear dependence of $S_V$ on temperature. Representative calibration curves are shown in the Supplementary Material section 1.

4. **QPC Transmission**

A small oscillating voltage $V_{ac}$ was sourced on source contact S at frequency $f_0$, corresponding to the tank circuit's center frequency (see Methods, *Noise acquisition setup*). The transmitted signal was collected at the amplifier ohmic contact (contact A in Fig. 1c of the main text) and measured with a spectrum analyzer with a resolution bandwidth of 100 Hz. To verify the absolute conductance scale, complementary low-frequency lock-in measurements were performed for both devices.

# Supplementary Materials

## 1. Thermal noise calibration

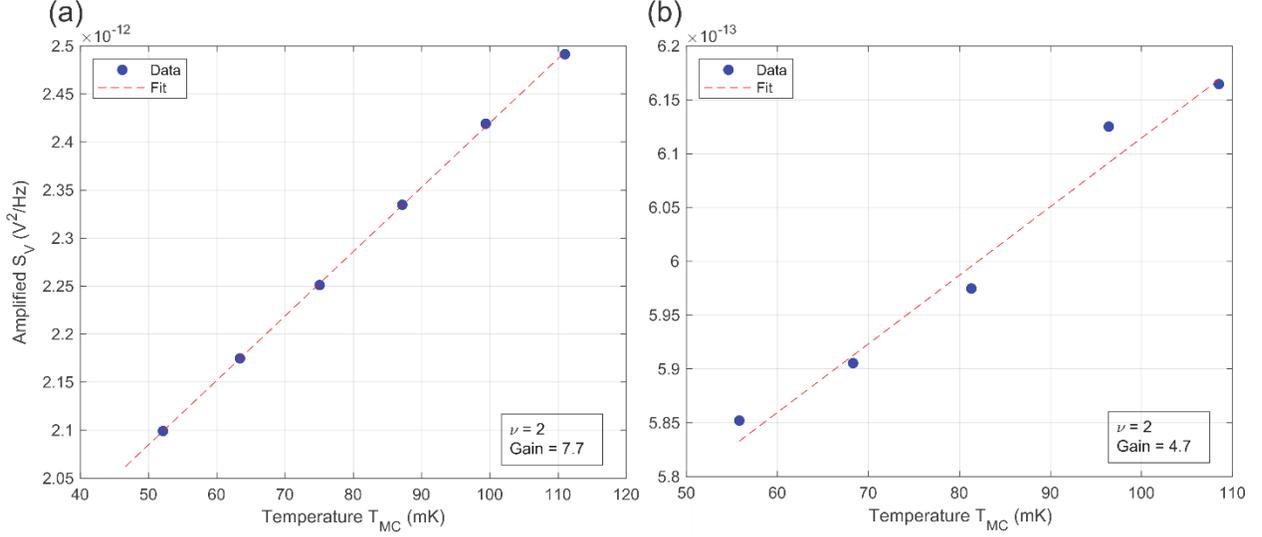

**Supplementary Figure S1:** Voltage noise spectral density measured as a function of the mixing chamber temperature $T_{MC}$. Panel (a) shows the data for device A, and panel (b) shows the corresponding data for device B. Both devices were calibrated at $\nu = 2$. See Methods, *Thermal noise gain calibration* for measurement details.

## 2. Crossover between thermal noise and shot noise

Shot noise due to a QPC is often analyzed using a phenomenological *independent-quasiparticle partitioning* expression, in which the thermal-to-shot-noise crossover is a function of the ratio $\frac{e^*V}{k_B T}$, and is given by:

$$S_I(0) = 2e^* I_{in} t(1-t) \left[ \coth\left(\frac{e^*V}{2k_B T}\right) - \frac{2k_B T}{e^*V} \right]. \tag{S1}$$

Here $e^*$ is the tunneling (partitioned) charge, $I_{in}$ is the injected DC current, and $t$ is the transmission coefficient of the QPC.

Recently, a theoretically motivated alternative has been proposed [1,2], in which the *shape* of the thermal-to-shot-noise crossover is sensitive not only to the quasiparticle charge but also to the scaling dimension $\Delta$ of the tunneling operator in chiral Luttinger Liquid (CLL) theory[3]. Within this framework, the low-frequency noise is given by,

$$S_I(0) = 2e^* I_{in} t(1-t) Im\left[\frac{2}{\pi}\psi\left(2\Delta + i\frac{e^*V}{2\pi k_B T}\right)\right], \qquad (S2)$$

where $\psi$ is the Digamma function, and $Im$ denotes the imaginary part.

For $\Delta = 1/2$, which corresponds to the expected scaling dimension of electron tunneling in noninteracting or integer quantum Hall edge modes, the CLL expression reduces to the phenomenological independent-quasiparticle form. However, for other values of $\Delta$, as expected for fractional quantum Hall states, the thermal-to-shot-noise crossover predicted by equation (S2) differs substantially from that described by (eq. S1). As a result, comparison of the measured crossover to the CLL-based expression provides a means to extract the scaling dimension $\Delta$.

This approach was subsequently implemented experimentally by Veillon *et al.* [4], who analyzed the thermal-to-shot-noise crossover at filling factors $\nu = 1/3$, $2/5$, and $2/3$. In those filling factors, the authors reported improved agreement with the CLL-based crossover expression, as evaluated using the theoretical values of $\Delta$, compared with the phenomenological form corresponding to $\Delta = 1/2$.

At $\nu = 1/2$, the scaling dimension $\Delta$ acquires particular significance, as different candidate topological orders predict distinct quasiparticle scaling dimensions. For example, the Moore–Read Pfaffian and the Halperin (331) states are characterized by different edge theories with corresponding tunneling exponents. Consequently, extending the CLL-based analysis of the thermal-to-shot-noise crossover to $\nu = 1/2$ offers, in principle, a route to distinguishing between competing topological orders via their predicted values of $\Delta$.

Using our $\nu = 1/2$ measurements, we performed a direct comparison between the independent-quasiparticle crossover formula used in the main text and the CLL crossover prediction evaluated with candidate $\nu = 1/2$ scaling dimensions.

Fig. S2 shows the Fano factor, defined as

$$F = \frac{S_I(0)}{2eI_{in}t(1-t)},$$

being plotted as a function of the normalized bias $\frac{eV}{k_B T}$. The measured crossover is compared with the independent-quasiparticle crossover that corresponds to $\Delta = 1/2$, as well as to the CLL predictions of the Pfaffian order ($\Delta = \frac{1}{8}$) and the 331 order ($\Delta = \frac{3}{16}$). For both devices, the crossover is well captured by the independent-quasiparticle form, further supporting the analysis presented in the main text.

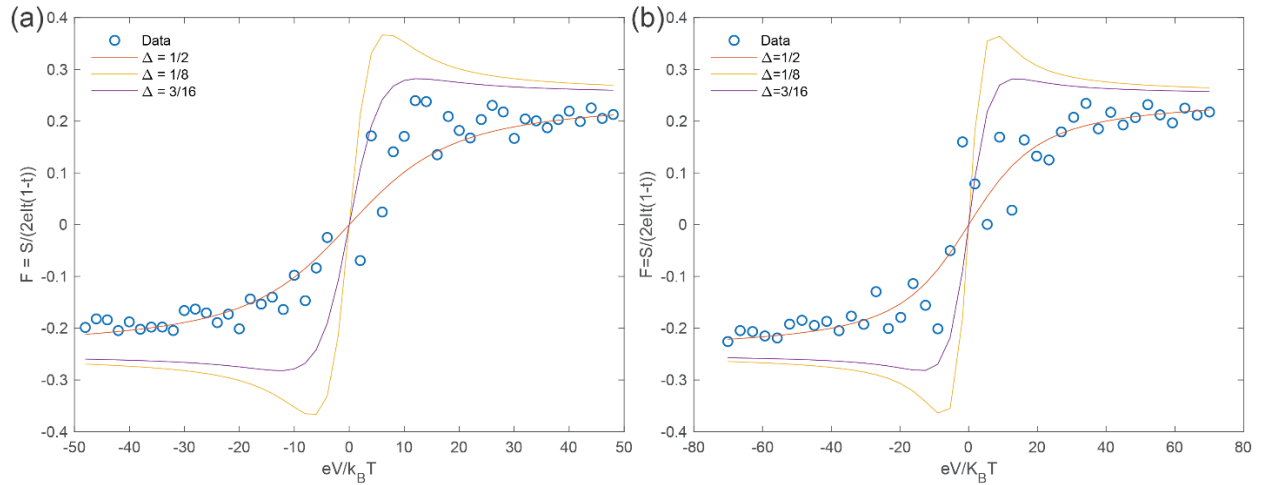

**Supplementary Figure S2: Crossover between thermal noise and shot noise.** Panels (a) and (b) show the measured Fano factor (blue dots) as a function of the normalized bias $eV/k_B T$ for devices A (Left) and B (Right), respectively. The red curve corresponds to the independent-quasiparticle partitioning model, which is equivalent to a scaling dimension $\Delta = 1/2$. The purple and yellow curves show the crossover predicted by the CLL expression (equation (S2)), using the theoretical scaling dimensions of the 'Halperin 331' state ($\Delta = 3/16$) and the Pfaffian state ($\Delta = 1/8$), respectively. In both devices, the measured crossover is well described by the $\Delta = 1/2$ model and deviates significantly from the predictions for $\Delta = 3/16$ and $\Delta = 1/8$.

Furthermore, this observation is consistent with the measured bias dependence of the QPC transmission. Over the bias range explored in these measurements, the transmission remains essentially constant, showing no dependence on the applied voltage. Within the CLL framework, deviations from $\Delta = 1/2$ are generally expected to be accompanied by nonlinear current–voltage characteristics. The absence of such effects in our data is therefore consistent with an effective scaling dimension close to $\Delta = 1/2$.

The absence of the strong nonlinear behavior predicted by ideal chiral Luttinger liquid theory has been reported previously in shot-noise and tunneling experiments at several fractional quantum Hall states [5]. This discrepancy between theoretical predictions and experimental observations has been widely discussed in the literature. It is commonly attributed to the renormalization of the effective scaling dimension governing quasiparticle tunneling at quantum point contacts [6]. Proposed mechanisms include interactions [7,8], coupling between multiple edge modes [9,10], $1/f$ noise [11], and the detailed shape and smoothness of the confining potential in the constriction region [12]. Such renormalization effects can drive the effective scaling dimension toward the noninteracting value $\Delta = 1/2$, thereby suppressing the expected nonlinear behavior and obscuring the intrinsic scaling dimension associated with the underlying topological order.

## 3. Beyond the weak backscattering regime

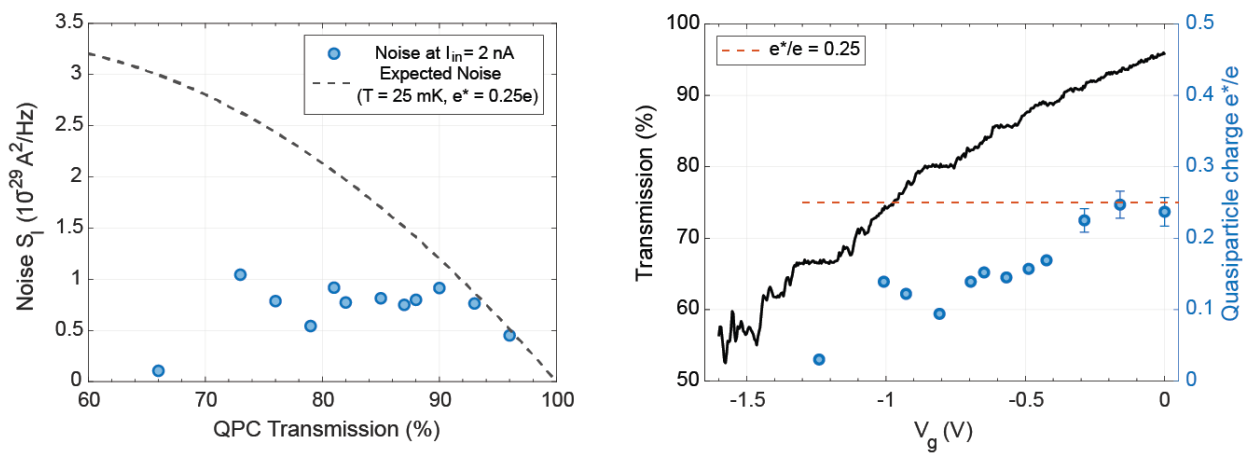

**Supplementary Figure S3:** Shot noise and effective charge beyond the weak-backscattering regime for device B. **Left:** Zero-frequency current noise measured at a fixed DC bias (2nA) as a function of QPC transmission. The noise does not follow the binomial t(1−t) dependence for e* = 0.25e, as expected for simple partitioning of a single edge mode with reduced transmission. **Right:** QPC transmission (black, left axis) and extracted quasiparticle charge from shot-noise measurements (blue, right axis) as a function of QPC gate voltage $V_g$. As the gate becomes more negative and the transmission decreases (below ~90%), the extracted quasiparticle charge deviates from the value obtained in the high-transmission regime. The orange line indicates e* = 0.25e.

As discussed in the main text, the stability of the ν=½ state depends sensitively on the local electrostatic environment[13]. As the QPC's gate voltage is increasingly more negative, reducing the transmission, it locally modifies both the filling factor, the electron density in the QPC, and the ν = ½ FQHS wavefunction asymmetry. Such changes are expected to alter the internal structure of the partitioned edge modes, potentially leading to 'edge reconstruction' and a breakdown of simple single-mode transport in the QPC.

Consistent with this picture, the measured shot noise as a function of the QPC's transmission does not follow the binomial expression t(1−t), which depends on independent tunneling events of quasiparticles. Consequently, the noise exhibits pronounced deviations, indicating that transport in this regime involves additional mechanisms, such as mode mixing, multiple tunneling paths, or suppression of the incompressible ν=1/2 phase within the QPC constriction. As a result, the effective charge extracted from shot-noise measurements in this regime cannot be straightforwardly interpreted as the fundamental quasiparticle charge of the bulk state.

Further evidence for this breakdown is provided by the behavior of the noise on conductance plateaus associated with fractional states such as ν=2/5 and ν=1/3. On these plateaus, the noise is strongly suppressed, and the extracted effective charges take unphysical values. In contrast to previous shot-noise measurements on well-developed plateaus, which typically yield the bulk quasiparticle charge (Fano factor of the bulk)[14-16]. This behavior suggests that the electronic structure within the QPC differs from that of the bulk, possibly due to edge reconstruction, thereby validating our approach for extracting the quasiparticle charge at high transmissions.

# 4. Charge distribution symmetry

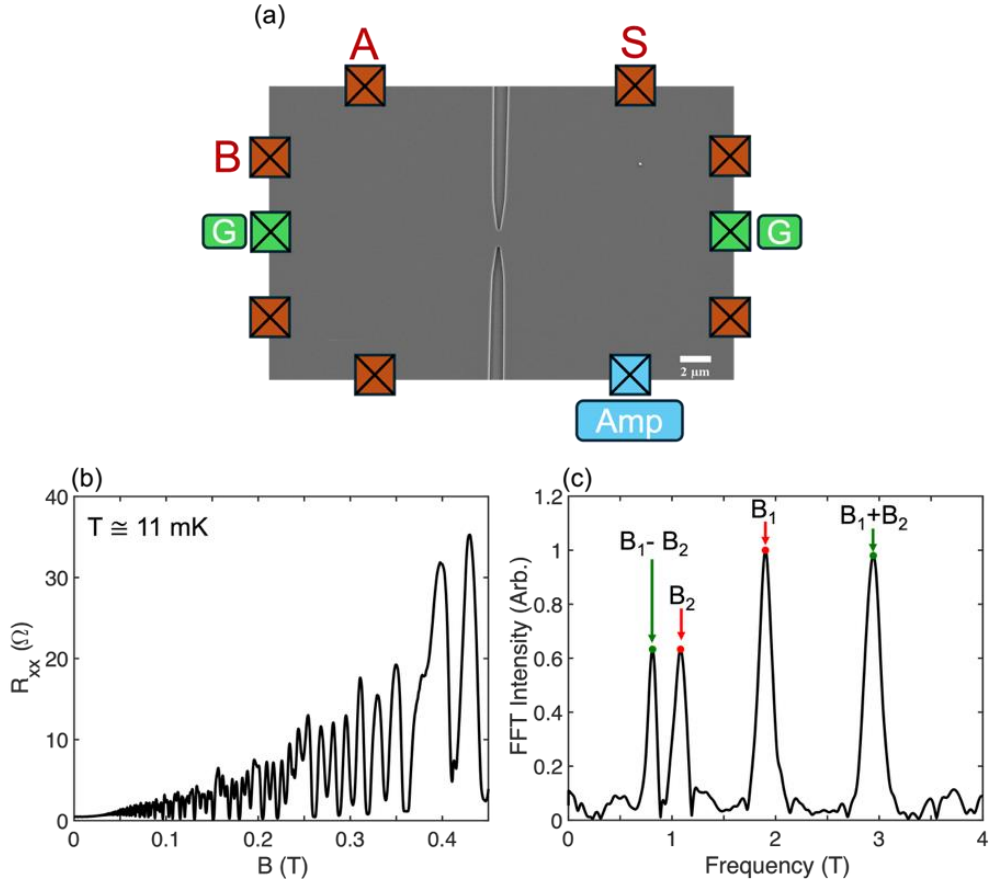

**Supplementary Figure S4:** Data measured in device A. (a) Sample schematic. (b) Low-field Shubnikov–de Haas (SdH) oscillations. For measurements of SdH oscillations, we passed current from contact S (source) to contact G (ground), and measured voltage difference between contact A and B. The electron temperature is 11 mK. (c) Fourier transform of the SdH oscillations. The red peaks marked as $B_1$ and $B_2$ correspond to the spin-unresolved frequencies of the lowest and the second-lowest electric subbands, respectively. The combination frequencies ($B_1 \pm B_2$) are also observed, as highlighted by green arrows.

The symmetry of the charge distribution is deduced from the energy separation ($\Delta_{12}$) between the lowest two electric subbands. This separation is determined from the fast Fourier transform (FFT) power spectrum of low-field Shubnikov–de Haas (SdH) oscillations, as shown in Fig. S4(b). Four distinct peaks are observed in the FFT spectrum as marked in Fig. S4(c): the peaks marked by red arrows correspond to the lowest ($B_1$) and second-lowest ($B_2$) electric subbands, whose carrier densities are given by $n_{1,2} = 2eB_{1,2}/h$, where the factor of 2 accounts for spin degeneracy. The peaks marked by green arrows arise from the combination frequencies. $B_1 \pm B_2$. From the extracted subband densities, $\Delta_{12}$ is obtained via

$\Delta_{12} = (n_1 - n_2)/(m^*/\pi\hbar^2)$, where $m^*$ is the effective electron mass. At a fixed density, $\Delta_{12}$ reaches its minimum for symmetric charge distribution and increases with an imbalance [17].

Comparing the experimentally extracted value of $\Delta_{12} \cong 16.4$ K to the results of Schrödinger–Poisson self-consistent simulations, as well as experimental data in similar samples, where we use both back and front gates to change the symmetry of the charge distribution and measure $\Delta_{12}$, we estimate a charge imbalance ($\Delta n/n$) of approximately 17% in our sample. ($\Delta n$ is defined by increasing $\Delta n/2$ from one gate and reducing $\Delta n/2$ from the other gate, resulting in a total difference of $\Delta n$ charge between the two interfaces of the quantum well.)